\documentclass[pdflatex,sn-mathphys-num]{sn-jnl}% Math and Physical Sciences Numbered Reference Style
\usepackage{graphicx}%
\usepackage{multirow}%
\usepackage{amsmath,amssymb,amsfonts}%
\usepackage{amsthm}%
\usepackage{mathrsfs}%
\usepackage[title]{appendix}%
\usepackage{xcolor}%
\usepackage{textcomp}%
\usepackage{manyfoot}%
\usepackage{booktabs}%
\usepackage{algorithm}%
\usepackage{algorithmicx}%
\usepackage{algpseudocode}%
\usepackage{listings}%
\theoremstyle{thmstyleone}%
\theoremstyle{thmstyletwo}%

\theoremstyle{thmstylethree}%

\begin{document}

\title[Article Title]{All-Optical Control over Nonlocality for Ultrafast Image Processing with an Excitonic Metasurface}


\author[1]{\fnm{Bernardo} \sur{S. Dias}}%\email{iauthor@gmail.com}
\equalcont{These authors contributed equally to this work.}

\author[2]{\fnm{Romain} \sur{Tirole}}%\email{iiauthor@gmail.com}
\equalcont{These authors contributed equally to this work.}

\author[2,3]{\fnm{Michele} \sur{Guizzardi}}%\email{iiiauthor@gmail.com}

\author[4]{\fnm{Andrea} \sur{Cordaro}}
\presentaddress{%
A. Cordaro: John A. Paulson School of Engineering and Applied Sciences,
Harvard University, 29 Oxford Street, Cambridge, MA, 02138, USA
}
%\presentaddress{John A. Paulson School of Engineering and Applied Sciences, Harvard University, 29 Oxford Street, Cambridge, MA, 02138, USA}%\email{iiiauthor@gmail.com}

\author[4]{\fnm{Albert} \sur{Polman}}%\email{iiiauthor@gmail.com}

\author*[2]{\fnm{Andrea} \sur{Alù}}\email{aalu@gc.cuny.edu}

\author*[1]{\fnm{Jorik} \sur{van de Groep}}\email{j.vandegroep@uva.nl}

\affil*[1]{\orgdiv{Van der Waals-Zeeman Institute, Institute of Physics}, \orgname{University of Amsterdam}, \orgaddress{\street{Science Park 904}, \city{Amsterdam}, \postcode{1098 XH}, \country{the Netherlands}}}

\affil[2]{\orgdiv{Photonics Initiative, Advanced Science Research Center}, \orgname{City University of New York}, \orgaddress{\street{85 St. Nicholas Terrace}, \city{New York}, \postcode{10031}, \country{USA}}}

\affil[3]{\orgdiv{Dipartimento di Fisica}, \orgname{Politecnico di Milano}, \orgaddress{\street{Piazza Leonardo da Vinci}, \city{Milano}, \postcode{20133}, \country{Italy}}}

\affil[4]{\orgdiv{Center for Nanophotonics}, \orgname{NWO-Institute AMOLF}, \orgaddress{\street{Science Park 104}, \city{Amsterdam}, \postcode{1098 XG}, \country{the Netherlands}}}

%\affil[5]{\orgdiv{John A. Paulson School of Engineering and Applied Sciences}, \orgname{Harvard University}, \orgaddress{\street{29 Oxford Street}, \city{Cambridge, MA}, \postcode{02138}, \country{USA}}}


\abstract{Image processing lies at the foundation of many modern technologies, such as augmented reality and autonomous driving, yet conventional digital approaches remain energy-intensive and limited in speed. Nonlocal metasurfaces—2D structures engineered at the nanoscale to support delocalized, dispersion engineered resonances—provide a fast, energy-efficient and ultrathin platform to perform image processing directly on the light path. Introducing tunability in this platform is an outstanding challenge, and would enable dynamic real-time control over the implemented processing operation, facilitating flexible integration into adaptive and multifunctional photonic architectures. Here, we demonstrate optically tunable edge detection at ultrafast speeds by integrating a dielectric nonlocal metasurface with multilayer WS$_2$, whose strong exciton-driven optical response enables dynamic control of the metasurface nonlocality at sub-ps speeds. Using resonant optical pumping of the A-exciton in WS$_2$, the metasurface transfer function is rapidly switched from edge detection to bright-field imaging by tuning its spatial nonlocality. Operating in the visible spectral range at a wavelength around 700 nm, the device shows ultrafast switching times and reaches an amplitude modulation depth of 11.5 dB for normal incident light. This approach provides a reconfigurable, ultrathin, all-optical platform for adaptive optical computing systems and highlights the potential of the highly nonlinear properties of 2D materials for active metasurfaces at ultrafast speeds.}




\maketitle

\section*{Introduction}%\label{sec1}

Image processing underpins a multitude of defining technologies of the digital age. From autonomous driving to AI-driven image classification and virtual reality, these applications are set to drive societal change by modifying both physical and digital interactions. At the same time, the rapidly growing associated data throughput and computational load place increasingly stringent demands on digital hardware, both in processing speed and energy efficiency\cite{ProblemsComputing}. As a consequence, the development of computational hardware is struggling to keep up with these demands, motivating the search for new paradigms for high-speed computation\cite{MooreLawEnd, NeuromorphicComputingReview, PhotonicComputingRev}. In this context, analog optical computation, enabled by recent advances in nanofabrication technologies, leverages easy-to-integrate optical metasurfaces to perform image processing operations directly on light without the need for electrical power\cite{MathOperationsMetamaterials}. Such optical computing metasurfaces have emerged as both an alternative and accelerator to digital image processing, offering the prospect of faster, energy-efficient operations with a microscale footprint\cite{CordaroMathOperations,ValentineFlatOptics2020, PhotonicsAIAccelerator, PhotonicsForAI, NonLocalDenoising, ValentineIncoherent}.

Nonlocal optical metasurfaces -- ensembles of nanoscale light scatterers with a collective optical response whose dispersion in frequency and momentum can be carefully tailored -- have become a key platform for analog optical computation. By supporting delocalized, dispersion-engineered optical modes, such nanopatterned ultrathin films encode mathematical operations directly in momentum space ($k$-space), avoiding the need for bulky 4$f$ systems of conventional Fourier optics\cite{vdLugtReview, ComputingReview} and the associated integration and alignment challenges. Notable demonstrations of this platform for image processing include edge detection \cite{NonlocalMetSigProc, CordaroMathOperations, DualPolImageProcessing, cotrufo2023dispersion, EdgeDetectReview}, phase-contrast imaging\cite{PhaseContrastMetasurface}, mathematical integration\cite{CordaroIntegration} and image denoising\cite{NonLocalDenoising}. Despite these advances, the applicability of such image-processing metasurfaces in optical computing to date has been hindered by their lack of reconfigurability: the designed mathematical operation is fixed at fabrication, constraining the device to a single image processing output only. Dynamic switching between different operations is therefore a crucial requirement for integration in real-world adaptive imaging systems. Achieving this feat requires active manipulation of a metasurface processing function, which has been recently demonstrated using phase-change materials\cite{ZhangPhaseChange, WangPhaseChange, CotrufoReconfigurable, PhaseChangeTemperature, PhaseChangeChalcogenide}, mechanical manipulation\cite{FaraonLens, ValentineStrain, PerspectiveMechanical,StrainHologram, MechanoOpticalMetasurfaces} and electrical gating\cite{LIDAR,RiseElectricTunMetasurf, ElectrSpaceTimeMetasurf, TomModulator, ElectricalModulation}, among others. Nevertheless, these approaches are limited to millisecond to second switching times, suffer from finite stability, and are challenging to integrate into existing systems, limiting their practical applicability.

Ultrafast optical modulation of a metasurface functionality has emerged as a promising route to overcome these limitations. Building on the nonlinear optical response of their constituent materials, pump-induced modulation has enabled dynamic switching of the metasurface optical response at ultrafast timescales\cite{CerulloNonLinearReview, TittlOpticalControl}, with demonstrations spanning resonance modulation, higher-harmonic generation, beam steering, and dynamic wavefront control\cite{TShegai3RMoS2, AltugMidIR_Modulation,AtwaterUltrafast}. However, nonlinear optical effects are generally weak in conventional materials, demanding high pump fluences that may be incompatible with practical device operation. Two-dimensional transition metal dichalcogenides (TMDCs) exhibit an exceptionally strong nonlinear optical response, which may be leveraged to address this limitation. In these semiconductors, efficient photon absorption induces nonlinear exciton-driven effects including the generation of an electron-hole plasma, exciton screening, and thermal bandgap renormalization (BGR), producing remarkably large and ultrafast variations in the refractive index (RI) \cite{Heinz_Mott, WS2MottDiscontinuous, CerulloMott,WS2CompactModulator}. While nonlinear TMDC metasurfaces have already been used for ultrafast modulation of second harmonic generation\cite{SHGMod, SHGModulator2, ZografTMDCNanoDisks, RahilZPL, ZotevTMDCReview}, their application for ultrafast control of a linear optical function encoded in its nonlocal response remains unexplored. 

Here, we combine the strong light-matter interaction of a tailored nonlocal metasurface with the strong nonlinear properties of a thin TMDC layer in a hybrid TMDC-metasurface device to demonstrate ultrafast, reconfigurable, all-optical image processing. By integrating a WS$_2$ flake on top of a nanobeam metasurface supporting a Fano resonance, we engineer the nonlocality of its response to perform switchable edge detection at 700 nm, and demonstrate sub-picosecond dynamic switching to bright-field imaging through optical pumping. Operating in both the CW and pulsed pump regime, we characterize the spectroscopic response of the device and report up to 11.5~dB amplitude modulation in sub-200 fs. Overall, our results demonstrate all-optical pumping of TMDC–metasurface hybrid devices as an ultrafast modulation mechanism for adaptive optical computing.

\section*{Results}%\label{sec1}
\subsection*{Concept of hybrid TMDC-metasurface for tunable optical computation}

\begin{figure}[h]
    \centering
    \includegraphics[width=\linewidth]{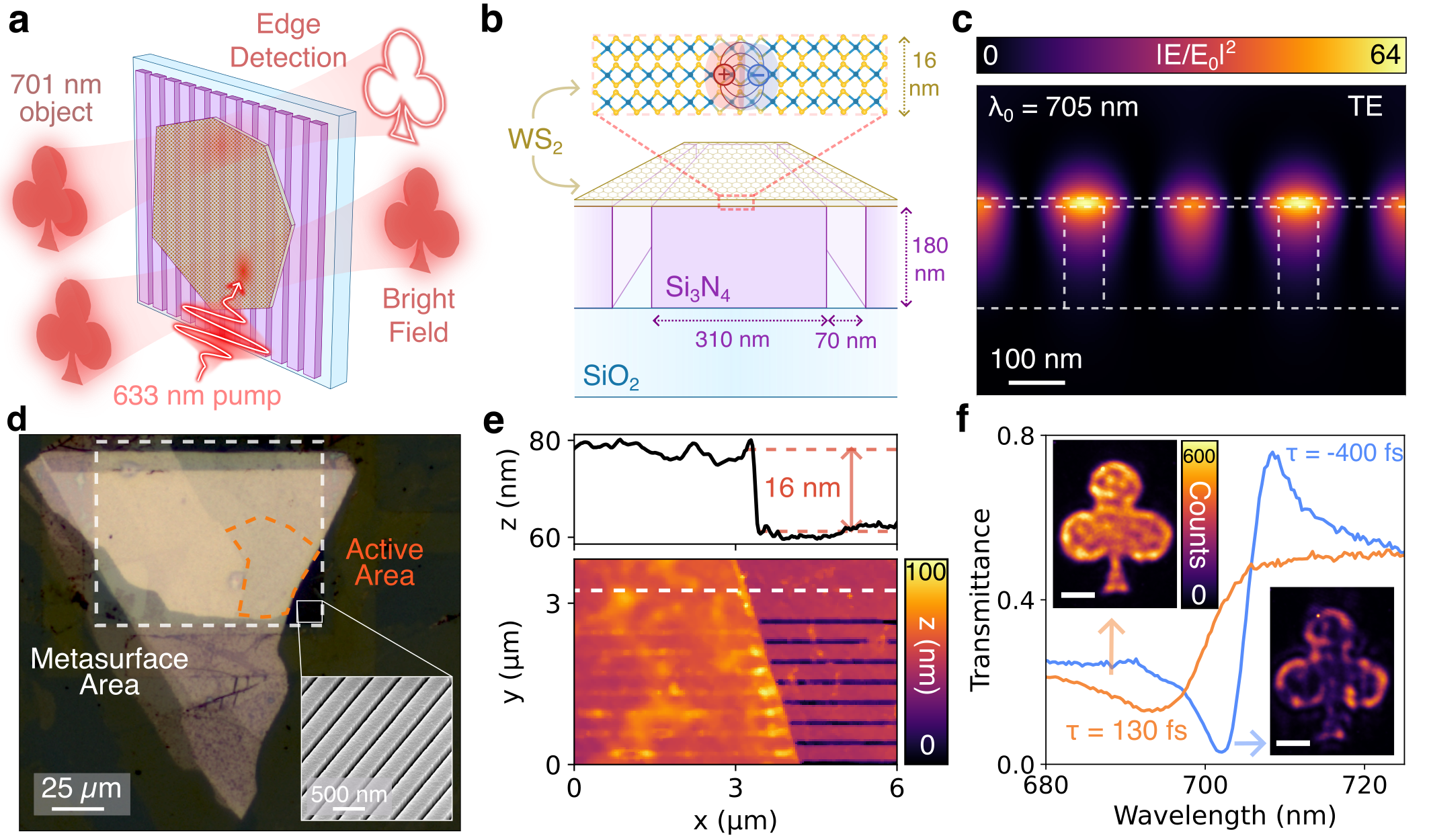}
    \caption{\textbf{Hybrid TMDC-metasurface device for metasurface image processing. (a)}~Schematic of the operation of the metasurface. In the absence of the optical pump (top), the nonlocal resonant response provides angular filtering leading to edge detection. Under optical pumping (bottom), the nonlocality is weakened, and a bright field image is recovered. \textbf{(b)} Schematic of the detailed geometry of the modulator design. Inset: 16 nm WS$_2$ layer showing an exciton. \textbf{(c)} Electric field intensity enhancement for a transverse electric (TE) polarized normal-incident wave at free-space wavelength $\lambda_0 = 705$~nm. The field is mostly confined to the WS$_2$ layer making the device response very sensitive to optical modulation of its RI. \textbf{(d)} Bright field microscope image of the device, showing the 2D material flake on top of the metasurface (inset: tilted SEM of bare metasurface). The active area with uniform WS$_{2}$ thickness is indicated in red. \textbf{(e)} Bottom: AFM scan of edge of the device, showing the 2D material on top of the metasurface. The white dashed line indicates the position of the cross section (top). \textbf{(f)} Spectral response of the metasurface under pumping, showing ultrafast switching from the edge detection (blue) to bright field imaging (orange). Insets: corresponding real-space images showing edge detection and bright field image at $\lambda=705$~nm, corresponding to the resonance modulation. Scale bar: 100 $\mu$m.}
    \label{fig:Concept}
\end{figure}

Figure~\ref{fig:Concept}a shows a schematic of the switchable image processing metasurface. To achieve tunable operation, we incorporate a 16~nm thin semiconducting WS$_2$ layer on top of a low-loss dielectric Si$_3$N$_4$ nanobeam metasurface. The nanopattern is comprised of 310~nm wide nanobeams with a thickness of 180~nm on a quartz substrate, and has a period of 380~nm (Fig. \ref{fig:Concept}b). It supports a guided mode resonance that interferes with a Fabry-P\'erot background in the far-field, giving rise to a high-$Q$ asymmetric Fano resonance at 705~nm, whose transmission minimum lies at 701 nm and serves as the edge-detection operating wavelength\cite{CordaroMathOperations}. The addition of the WS$_2$ layer introduces a second excitonic material resonance at 633 nm, whose pump-induced dynamics modify the refractive index (RI) of the layer. This in turn allows modulation of the nonlocality of the resonant response, enabling switching between edge detection and bright-field imaging (Fig. \ref{fig:Concept}a).

The Fano resonance plays two crucial roles. First, its nonlocal nature provides the $k$-space filtering required for edge detection: in the absence of pumping, the parabolic transfer function performs second-order spatial differentiation\cite{CordaroMathOperations}. Upon pumping, the nonlocality is weakened, and the image processing transfer function is abruptly modified. Second, its strongly confined resonant near fields and long interaction length with the WS$_2$ layer make the device response highly sensitive to exciton-driven RI variations (Fig. \ref{fig:Concept}c). The two resonances are intentionally spectrally separated to isolate optical pumping and the operation wavelength. The Fano resonance is designed below the WS$_2$ bandgap energy where the material's absorption is limited to weak defect-induced dissipation only\cite{TemperatureEllipsometryWS2, WS2CompactModulator}. This is a necessary condition for edge detection, as the optical resonance must reach near-zero transmission, which is incompatible with the large absorption present near the exciton wavelength. While the pump resonantly excites the exciton transition at 633 nm, the exciton's long-wavelength tail extends beyond 700 nm yielding sufficiently large pump-induced RI changes at 705 nm to dynamically tune the optical resonance\cite{WS2CompactModulator}.

We fabricate the metasurface using conventional electron-beam lithography and reactive-ion etching (see Methods), and integrate the WS$_2$ layer afterwards on top of the as-fabricated metasurface via a polymer-assisted dry transfer method that is particularly designed for stamping on patterned surfaces\cite{LDPE_Stamp} (Fig. \ref{fig:Concept}d). This workflow mitigates any fabrication-induced 2D material degradation. At the same time, the high quality factor of the resonance and large refractive index of the WS$_2$  make the resonance very sensitive to small local variations in WS$_2$ thickness and fabrication imperfections. As a result, we use a limited region of approximately 25 × 50 $\mu$m of the WS$_2$ flake with homogeneous thickness for edge detection (Fig. \ref{fig:Concept}d). Using atomic force microscopy (AFM) at the edge of the active area, we characterize the thickness of the WS$_2$ as 16 nm (Fig. \ref{fig:Concept}e). 

The metasurface switches between edge detection and bright-field imaging in sub-picosecond timescales (Fig. \ref{fig:Concept}f). In the absence of pumping, the Fano resonance produces a sharp dip in transmission reaching near-zero, yielding strongly delocalized response and enhanced signals at the edges compared to the bulk of the clover target. After the arrival of the pump on the other hand, the resonance dip is blueshifted and dampened, resulting in a local, momentum-independent response and thereby a bright-field image of the clover.

\subsection*{Continuous-wave optical modulation and edge detection}%\label{sec2}

\begin{figure}[h]
    \centering
    \includegraphics[width=\linewidth]{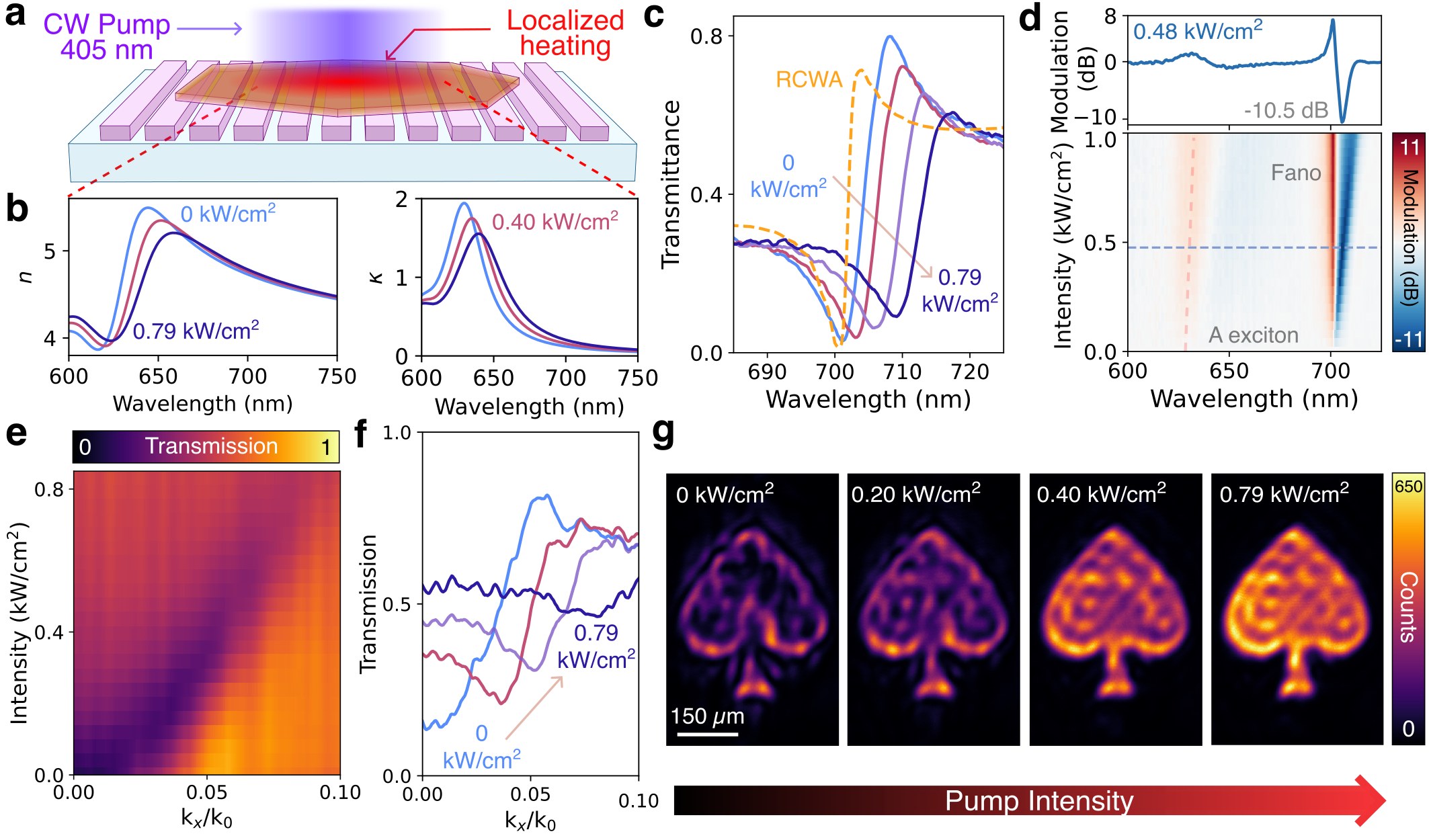}
    \caption{\textbf{Tunability of the hybrid TMDC-metasurface device under CW pumping at 405~nm. (a)} Schematic of the modulation mechanism under CW pumping. The WS$_2$ layer efficiently absorbs the pump, leading to localized heating. \textbf{(b)} Changes in the WS$_2$ RI under CW pumping, extracted from reflectivity of a bare flake on Si$_3$N$_4$. Under optical pumping, the exciton redshifts, leading to an increase of both real and imaginary RI at 705 nm. \textbf{(c)} Transmittance measurement of the metasurface under normal incidence for pump intensities of 0 (light blue), 0.20 (red), 0.52 (purple), and 0.79~kW/cm$^2$ (dark blue). The spectrum simulated using rigorous-coupled wave analysis (RCWA) is also shown (yellow-dashed). \textbf{(d) Bottom}: modulation of the resonance at normal incidence as function of pump intensity. Dashed blue line indicates cross section shown in top. \textbf{Top}: modulation spectrum at the intensity of maximum modulation (0.48 kW/cm$^{2}$), reaching -10.5dB. \textbf{(e)} Angular response of the transmission $|t(k)|$ as function of pump power, measured at the operation wavelength of 701~nm. \textbf{(f)} Transmission curves of the angular response of the device (cross cuts of (e) at the same pump fluences as in (c)), highlighting the switch in nonlocal response, from a parabolic filter to uniform transmission. \textbf{(g)} Tunable image processing of a chromium target projected onto the metasurface, showing a transition from edge detection to bright field imaging under increasing pump intensity. Note that the color bar and illumination conditions at 701~nm are identical for all four images.}
    \label{fig:CW_Modulation}
\end{figure}

To assess the tunable response of our device, we start by studying its transmittance under CW illumination with varying pump intensities (Fig. \ref{fig:CW_Modulation}a-c). We choose a pump wavelength of 405 nm to match the maximum of the WS$_2$ absorption spectrum\cite{WS2_Munkhbat} and use a uniform circular pump spot with 125 $\mu$m diameter that is larger than the active area where an image would be projected for edge detection (see Supplementary Information for setup description).

We first characterize the pump-induced changes to the RI of the bare WS$_2$ in absence of the metasurface. To this end, we measure the transmittance spectrum of the section of the WS$_2$ flake that is placed on the flat Si$_3$N$_4$ layer, outside of the nanopatterned region (full RI retrieval method in Supplementary Information). Under pumping, the A exciton shows a noticeable redshift that is accompanied by spectral broadening (Fig. \ref{fig:CW_Modulation}b). Its underlying mechanism is established as a combination of carrier-induced and thermal BGR, both contributing to an overall redshift\cite{WS2CompactModulator, TemperatureEllipsometryWS2, CW_Modulation1, CW_Modulation2}. At the resonance wavelength of 705 nm, this redshift results in an increase of both the real and imaginary parts of the WS$_2$ RI. Indeed, also the TMDC-metasurface device (Fig. \ref{fig:CW_Modulation}c) exhibits a redshift and increased damping in its resonant response under normal incidence, with no hysteresis under repetitive cycling of the optical pump (see Supplementary Material). As the pump power increases, the resonance minimum redshifts progressively and, at 0.48 kW/cm$^2$, crosses the unpumped transmission maximum at 708 nm, yielding a maximum modulation depth of $10 log_{10}(I_{P=0.48}/I_{P=0})=-10.5$~dB (Fig. \ref{fig:CW_Modulation}d). %We note that in the absence of pumping, at the edge detection wavelength of 701 nm, the resonance dips down to 2.5\%, with the residual WS$_2$ absorption at this wavelength preventing the resonance from reaching zero. 
 The $\sim30$\% absolute contrast in transmission between pumped and unpumped states suggests that switching between edge detection and bright-field imaging is feasible.

Edge detection using metasurfaces requires efficient filtering of the low-$k$ components of an image while transmitting the high-$k$ components that result from scattering off an object's edges\cite{CordaroMathOperations}. Crucially, the increased absorption under pumping weakens the nonlocality of the mode, which affects its ability to perform edge detection. To evaluate the angular response of our device as a function of pump intensity, we measure metasurface's transmission $|t(k)|$ obtained via back focal plane (BFP) imaging at the minimum wavelength of 701 nm (Fig. \ref{fig:CW_Modulation}e,f). We observe that the spectral redshift with increasing pump intensity (Fig. \ref{fig:CW_Modulation}c) pushes the resonance to larger wave vectors, tracing a near-diagonal line in Fig. \ref{fig:CW_Modulation}e. As a result, the angular transmission curve $|t(k)|$ evolves from a parabolic profile in the unpumped state (i.e. strong angular filtering) to a near uniform response with $|t|\sim0.55$ under pumping with $0.79$~kW/cm$^{2}$ (Fig. \ref{fig:CW_Modulation}f). We note that in Fig. \ref{fig:CW_Modulation}c the resonance minimum dips down to the non-zero value of $T=2.5$\% as a result of residual defect-induced absorption in the WS$_2$, even at photon energies lower than the band gap. This causes the parabolic transmission response to achieve a minimum of $|t|=\sqrt{T}=16$\% in transmission. Nevertheless, such level of residual transmission does not compromise the edge detection functionality\cite{CordaroMathOperations}, and indeed a clear and continuous switch from edge detection to bright-field imaging is observed with increasing the pump power (Fig. \ref{fig:CW_Modulation}g). We note that the 1D nature of the operation can be recognized from the bottom of the 'ace',  where no vertical line is observed in edge detection mode. Yet, all non-orthogonal edges are clearly observed. Overall, these results highlight how the hybrid 2D-metasurface offers remarkably strong tunability under CW illumination that can be used for light modulation and tunable image processing.

\subsection*{Ultrafast switchable nonlocality of excitonic metasurface}%\label{subsec3}

\begin{figure}[h]
    \centering
    \includegraphics[width=\linewidth]{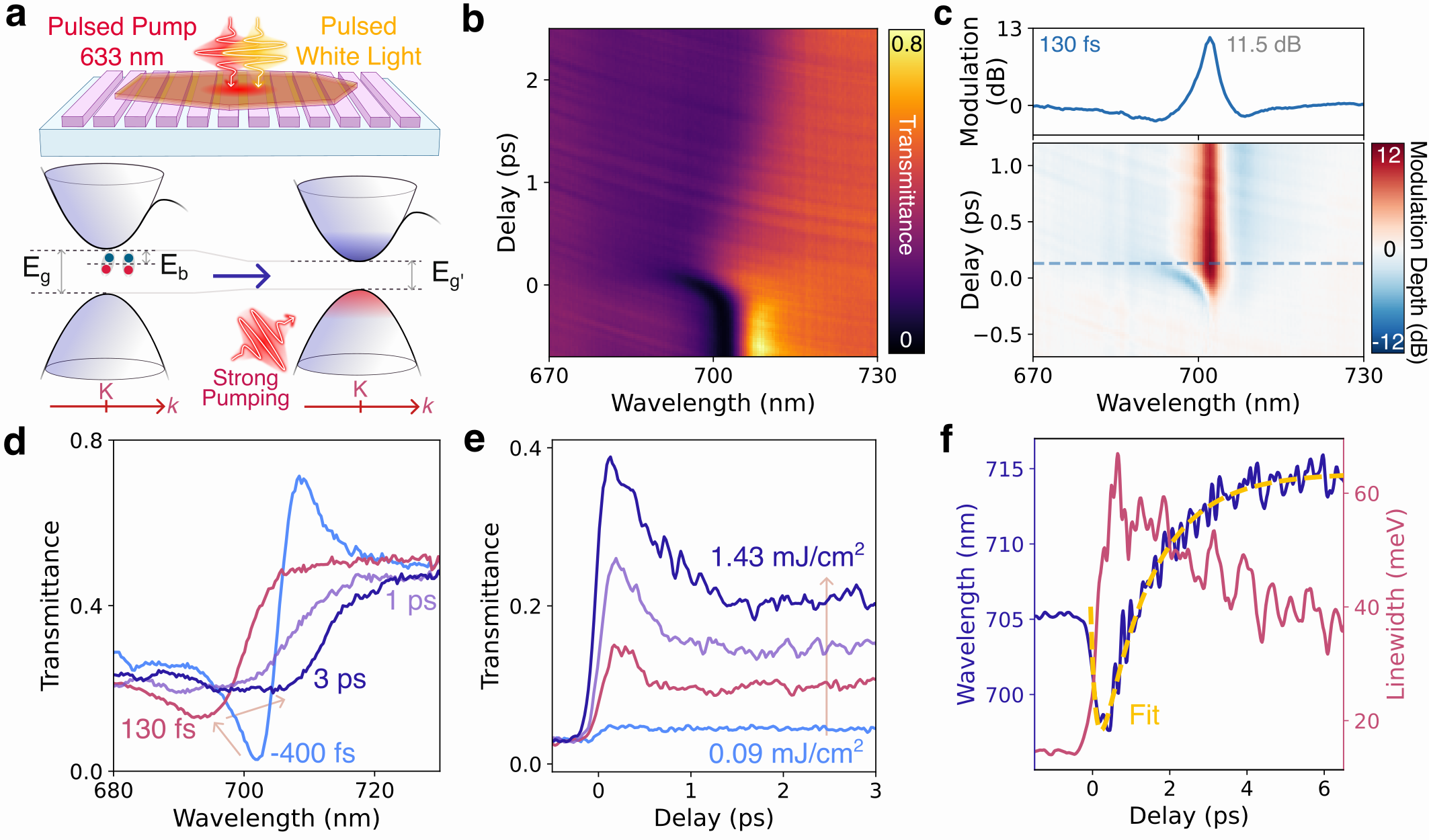}
    \caption{\textbf{Ultrafast metasurface operation under pulsed laser pumping at 633 nm. (a)}~Schematic of the modulation mechanism under pulsed laser pumping resonant with the A-exciton. The WS$_2$ layer absorbs the pump photons, leading to both increased screening and plasma-induced BGR. \textbf{(b)} Transmittance of the metasurfaces as a function of time-delay. \textbf{(c) Bottom}: time evolution of the modulation around the optical resonance. The horizontal blue-dashed line indicates the delay corresponding to the maximum modulation used in the top. \textbf{Top}: maximum modulation achieved by the device at $\Delta t=130$~fs, showing a modulation depth of 11.5 dB at $\lambda=701$~nm. \textbf{(d)} Transmittance under normal incidence for several delay times, showing the time evolution of the resonance. \textbf{(e)}~Time evolution of the transmission at normal incidence, measured at the resonance minimum of 701 nm for pump fluences of 0.17 (light blue), 0.39 (red), 0.66 (purple), and 1.43 mJ/cm$^2$ (dark blue). \textbf{(f)} Time evolution of the wavelength (blue) and linewidth (red) of the Fano resonance, showing the initial screening-induced blueshift and subsequent plasma-induced redshift. At later delays, the resonance stabilizes into a long-lived redshift caused by thermal effects. Yellow-dashed line shows a fit to the spectral evolution of the resonance.}
    \label{fig:Pulsed_Modulation}
\end{figure}

Excited-state carrier dynamics in TMDC materials unfold at the femtosecond timescale, well before thermal effects in the lattice become relevant\cite{TrovatelloExciton}. While the CW response characterized above represents the steady-state dominated by pump-induced heating, pulsed optical excitation offers the characterization of the metasurface at these ultrafast time scales (Fig. \ref{fig:Pulsed_Modulation}). %We now study the device response under pulsed laser excitation, in metasurface pump-probe experiments (Fig. \ref{fig:Pulsed_Modulation}). 
By resonantly pumping the WS$_2$ A-exciton at 633 nm, carriers are injected directly into the exciton state, bypassing any hot-carrier intra-band relaxation to the conduction band minimum (see Supplementary Materials for 405 nm pump-probe spectroscopy). This results in faster modulation speeds compared to above-band excitation\cite{CandDexcitonPumpProbeWS2}. 

We use cross-polarized 120 fs pulses for the pump and white light probe, and perform our experiments at a maximum single-pulse fluence of 1.43 mJ/cm$^{2}$, driving the WS$_{2}$ above the excitonic Mott transition threshold. In this regime the photoexcited exciton density is sufficiently large that exciton-exciton Coulomb screening overcomes the binding energy, ionizing the excitonic population and driving the material into a dense electron-hole plasma\cite{Heinz_Mott, WS2MottDiscontinuous, CerulloMott} (Fig. \ref{fig:Pulsed_Modulation}a). Two competing effects govern the optical response near the A-exciton resonance. On one hand, phase-space filling (Pauli blocking) and a reduced exciton binding energy due to screening give rise to a blueshift of the exciton transition. On the other hand, the dense plasma causes exciton bleaching and drives carrier-induced BGR, redshifting the band edge (E$_{g'}<$ E$_{g}$) and extending absorption to wavelengths below the original band gap. The relative weight of these contributions varies with excitation fluence, determining the net spectral response\cite{WS2MottDiscontinuous}.

To analyze the effect of resonant pumping on the metasurface optical response, we perform pump-probe measurements of its transmittance (Fig. \ref{fig:Pulsed_Modulation}b). In the first 150~fs following excitation, we observe a pump-induced blueshift of the resonance, along with strong damping. As the carriers thermalize, the resonance redshifts to $\lambda=714$~nm, resulting in a long-lived modification of the device transmittance due to thermal heating (see section 4 of the Supplementary Information). As a direct consequence of these dynamics, the initial blueshift induces an ultrafast ($\tau$ = 130 fs) increase in transmittance of approximately 35\% at the largest fluence, corresponding to a modulation of 11.5 dB at the resonance minimum (Fig. \ref{fig:Pulsed_Modulation}c). This large modulation is aided by the strongly asymmetric line shape of the Fano resonance that gives rise to a steep increase in transmittance at $\lambda=701$~nm as the resonance blueshifts (Fig. \ref{fig:Pulsed_Modulation}d,e). The subsequent redshift moves the resonance in the opposite direction, along the shallow flank, yielding a smaller modulation of 25\% in transmittance. The weak modulation observed at negative delays is attributed to the cavity ring-down of the high-$Q$ Fano resonance, which extends the probe response beyond the pulse duration\cite{FalcoProbeExtended}.

While the transmittance modulation is fluence-dependent, we observe similar trends for all values (Fig. \ref{fig:Pulsed_Modulation}e). From the analysis of a longer scan, we find that the device recovers to 90\% of its initial transmission within 396 $\pm$ 23 ps, in agreement with previous work\cite{ThermalRecoveryWS2, WS2GrapheneRelaxation, WS2CompactModulator}. This recovery time corresponds to an operational bandwidth of 2.5 GHz, orders of magnitude faster than a typical imaging camera. The bandwidth is limited by the thermal BGR-induced redshift of the excitonic resonance through lattice heating. For repetition rates exceeding $\sim$2.5 GHz, successive pulses arrive before the lattice has fully cooled representing a transition into the steady-state tuning in the CW regime.

To further quantify the TMDC-governed resonance dynamics, we fit a Fano line shape to spectra for all delay times and extract the resonance wavelength and linewidth (Fig. \ref{fig:Pulsed_Modulation}f). This analysis reveals that the initial blueshift is 6 nm, followed by a redshift of 16 nm. At the same time, the damping increases the linewidth from 15 meV to $\sim60$~meV at the moment of maximum blueshift. To relate the spectral shifts to fundamental processes in the material's carrier dynamics, we fit the extracted resonance energy shift with a bi-exponential model\cite{WS2MottDiscontinuous} (Fig. \ref{fig:Pulsed_Modulation}f):

$$\Delta E = R_{plasma}e^{-t/\tau_{plasma}} + B_{pop}e^{-t/\tau_{pop}} + C_{heat},$$

where $R_{plasma}$ is the electron–hole plasma redshift amplitude, $B_{pop}$ is the Pauli blocking blueshift amplitude, $\tau_{plasma}$ and $\tau_{pop}$ are the corresponding decay time constants, and $C_{heat}$ is a constant offset accounting for the long-lived thermal redshift. From this model, we extract $R_{plasma}$ = -57~$\pm$~5 meV and $\tau_{plasma}$ = 0.64~$\pm$~0.04 ps, $B_{pop}$ = 82~$\pm$~5 meV and $\tau_{pop}$ = 1.77~$\pm$~0.05 ps, and $C_{heat}$ = -26.57~$\pm$~0.07 meV, in agreement with previously reported A-exciton dynamics in WS$_2$\cite{WS2MottDiscontinuous}. This corroborates that the dynamics of the optical resonance are directly caused by pump-induced variations in the WS$_2$ refractive index: Pauli blocking induces an initial blueshift of the A-exciton resonance, while simultaneous electron–hole plasma formation produces a competing redshift via BGR. Both contributions decay on sub-picosecond to few-picosecond timescales, leaving a long-lived thermal redshift. Both mechanisms result in a smaller but concomitant change in the RI around $\lambda=705$~nm.
By virtue of the strongly enhanced light-matter interaction of the metasurface, the pump-induced blueshift and simultaneous linewidth broadening suppress the nonlocal Fano response of the device, leading to a 35\% increase in absolute transmittance, enabling switching between edge detection and bright field imaging.

\begin{figure}[ht!]
    \centering
    \includegraphics[width=\linewidth]{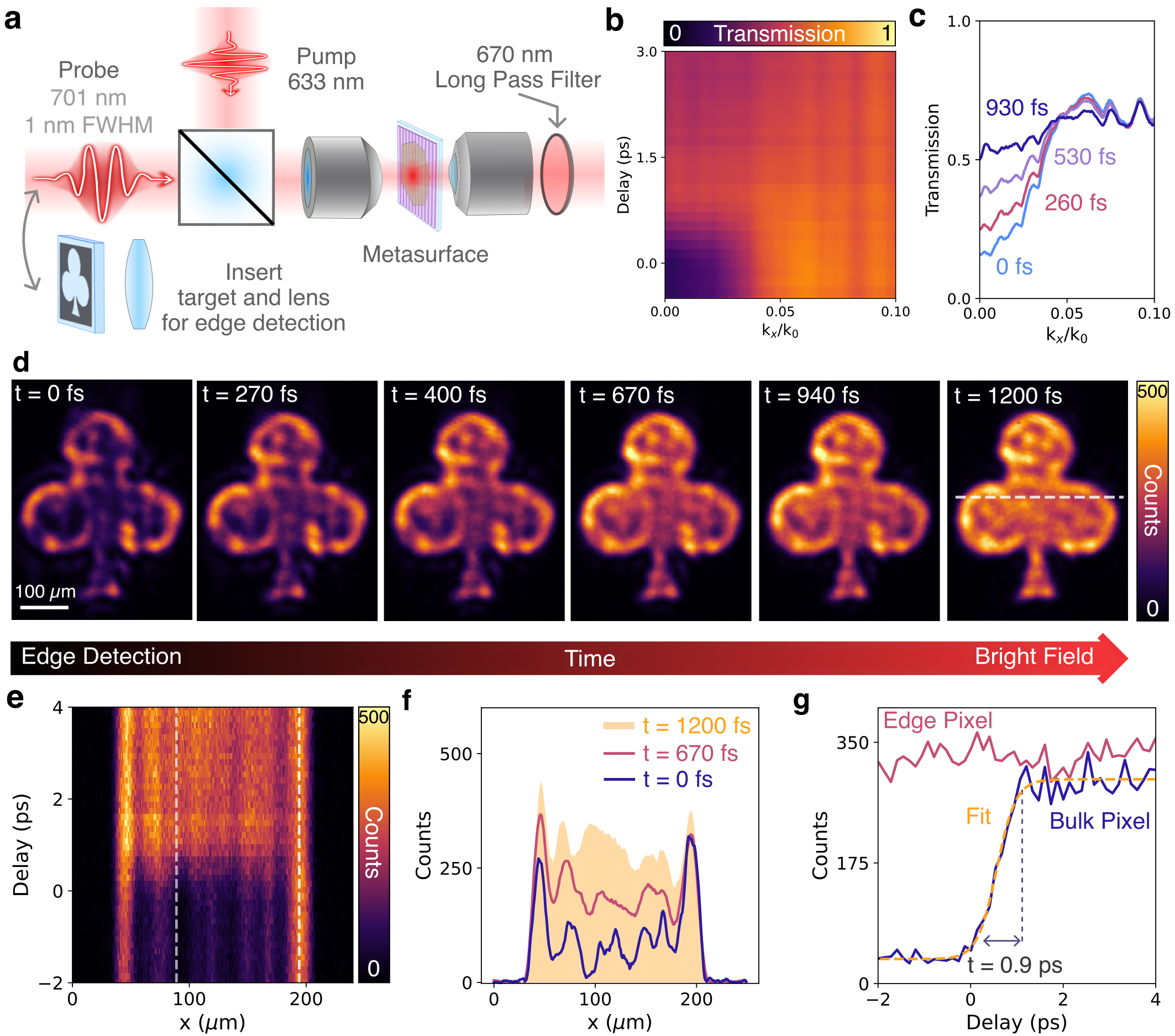}
    \caption{\textbf{Ultrafast control of image processing under pulsed laser illumination at 633 nm. (a)} Experimental setup used for BFP and edge detection measurements. \textbf{(b)} Time evolution of the angular response $|t(k)|$ of the device under pumping, showing ultrafast damping of the parabolic response leading to uniform transmission. \textbf{(c)} Transmission curves at multiple delay times (cross cuts of (b)), highlighting the transition from a parabolic (edge detection) to uniform (bright field) transmission. \textbf{(d)} Ultrafast tunable image processing of a chromium target projected on the metasurface, showing transition from edge detection to bright field as function of time. The dashed white line indicates the position of the line section used in (e). \textbf{(e)} Time evolution of the line section showing that the inside of the image transitions from dark to bright, while the edges keep approximately the same brightness. The vertical lines indicate the center pixels used in the time traces in (g). \textbf{(f)} Line section of (e) before and after optical pumping, emphasizing the transition from edge detection to bright field transmission. \textbf{(g)} Time evolution of a pixel on the edge of the image and on the inside, showing the sub-picosecond temporal evolution of the bulk of the image while the edges remain constant.}
    \label{fig:Pulsed_EdgeDetection}
\end{figure}

Capitalizing on the metasurface ultrafast response, we now characterize its image processing performance under pulsed laser illumination (Fig. \ref{fig:Pulsed_EdgeDetection}). To achieve optimal edge detection, we tune the probe to the minimum of the Fano resonance ($\lambda=701$~nm) and reduce its bandwidth to 1 nm FWHM (Fig. \ref{fig:Pulsed_EdgeDetection}a). While this improves edge detection images, it also lengthens the probe pulse in the temporal domain ($\tau_{min} \approx0.73$ ps), leading to lower time resolution speeds than observed in Fig. \ref{fig:Pulsed_Modulation}f. First, to characterize the angular response of the device, we analyze the BFP response at a fluence of $1.43$~mJ/cm$^{2}$ (Figs. \ref{fig:Pulsed_EdgeDetection}b,c). We observe that the parabolic $k$-space response $|t(k)|$ transforms into a uniform, near-flat, transmission of $t=0.55-0.6$ in less than 1 ps, enabling a fast transition from edge detection to bright field imaging. Given the spectral dynamics presented in Fig. \ref{fig:Pulsed_Modulation}b–f, we verify that the pump-induced wavelength shift and linewidth broadening of the Fano resonance collectively suppress its momentum-selective nonlocal response, driving the angular transfer function from a parabolic profile toward a spectrally flat, local transmission.

As a final assessment of its switchable performance, we project the image of a clover on the metasurface and analyze the time evolution in real-space as a function of pump-probe delay (Fig. \ref{fig:Pulsed_EdgeDetection}d). At $\tau \leq$ 0 fs, we observe an image with enhanced edges, quickly transitioning to a bright field version upon arrival of the pump. Taking a cross section of the image shows that the time evolution of the bulk of the image displays a clear transition from dark to bright, while the edges show an approximately constant response (Fig. \ref{fig:Pulsed_EdgeDetection}e,f). This is in agreement with the BFP response, where the optical response for high frequency components shows little variation as a function of pump modulation (Figs. \ref{fig:Pulsed_EdgeDetection}b,c). To determine the switching speed of the operation, we compare the time evolution of a pixel corresponding to the bulk of the image with one on the edge of the image (Fig. \ref{fig:Pulsed_EdgeDetection}g). While the edge pixel remains constant after pump excitation, the bulk pixel increases in intensity by a factor 6 within a time constant of $0.91\pm0.09$~ps, enabling the metasurface function to switch between edge detection and bright field imaging. We note that in our experiment the observed temporal response is limited by the spectrally filtered probe, whose duration is Fourier-limited by the 1~nm spectral bandwidth imposed by the high-$Q$ resonance. Full access to the sub-200 fs dynamics observed in Fig. \ref{fig:Pulsed_Modulation}b-f could be achieved by filtering the probe after interaction with the sample. Overall, these results demonstrate ultrafast optically reconfigurable nonlocality of our hybrid TMDC-metasurface device, achieving sub-ps control of image processing, switching between edge detection and bright-field imaging modes driven by excitonic pump-induced changes to the nonlocality of its Fano resonance.

\section*{Conclusions}

By harnessing the interplay between optical and excitonic resonances in a hybrid TMDC-nonlocal metasurface, we demonstrate reconfigurable all-optical image processing with sub-picosecond switching times. Capitalizing on the nonlocality and field enhancement of the metasurface and the strong pump-dependent RI of the TMDC, we show ultrafast 11.5 dB optical modulation and continuous tuning from edge detection to bright-field imaging. While the present design is limited to edge detection along a single spatial direction, the approach is readily extendable to 2D edge detection using an isotropic angular transfer function\cite{ValentineFlatOptics2020,DualPolImageProcessing}. Beyond spatial filtering, the ultrafast timescales demonstrated here open perspectives for devices that operate simultaneously in the spatial and temporal domains, where engineered photonic–excitonic responses could enable computation through joint space–time filtering. At a higher level, the strong field confinement at the thin TMDC layer makes this hybrid metasurface a promising platform to study fundamental ultrafast light–matter interactions in 2D materials, including nonlinear excitonic dynamics and carrier–photon coupling at femtosecond timescales. 

\section*{Methods}
\subsection*{Optical Device Fabrication}

We fabricate the metasurface by deposition of 180 nm of Si$_3$N$_4$ on a clean SiO$_2$ substrate via inductively coupled plasma chemical vapor deposition (Oxford PlasmaPro100 ICPECVD). For nanobeam patterning, we employ electron-beam lithography using a CSAR resist layer (AR-P 6200.09). After development, the Si$_3$N$_4$ is etched using a CHF$_3$/O$_2$ plasma (55 and 5 sccm, respectively) at a forward power of 175 W (Oxford Plasmalab 80+). Finally, the residual resist is removed by immersion in anisole at 60$^{\circ}$C.

The 2D material flake is obtained via Au-assisted exfoliation of bulk WS$_2$\cite{AuExfol} on SiO$_2$. While this method produces mostly TMDC monolayers, some thicker flakes can also be obtained, generally with larger areas than with mechanically exfoliated techniques. The thickness of the flake was measured via optical reflectometry and transfer matrix model fitting. The WS$_2$ is transferred using the LDPE-based dry-transfer technique\cite{LDPE_Stamp}, followed by oleic acid cleaning and vacuum annealing at 180$^{\circ}$C. 

\subsection*{Sample Characterization}

For the AFM measurements, a Bruker Dimension FastScan system is used at a rate of 13 \textmu m/s. The SEM images are taken in a FEI Verios 460 system.

\subsection*{Numerical Simulations}

Numerical simulations are performed using the Python version of the S$^4$ RCWA package\cite{S4}. In these simulations, we obtain the RI of SiO$_2$ and WS$_2$ from literature as tabulated data\cite{SiO2_Malitson, WS2_Munkhbat}. For Si$_3$N$_4$, we measure the RI through spectroscopic ellipsometry (J.A. Woollam VB-400) of the material on a Si substrate.

\subsection*{Spectroscopy and Imaging Measurements}

Both spectroscopic and edge detection measurements are made in a custom-built microscopy setup that allows optical pumping with both CW and pulsed lasers (see Supplementary Materials, section 1). For the CW measurements, the output of a supercontinuum laser (NKT SuperK Fianium FIU-15) is passed through a pulse shaper, enabling the selection of wavelengths with 1 nm FWHM resolution. In this case, a computer controlled 1W 405 nm CW laser (Axiom Optics Lambda Mini) is used as pump and is inserted in the setup via a multimode optical fiber (Thorlabs M36L01). The pump-probe experiments are carried out using 120 fs (FWHM intensity) duration pulses from a Coherent Astrella laser with a 1 kHz repetition rate. The pump and probe pulses frequency are respectively tuned using LightConversion Opera and Topas optical parametric amplifiers. A motorised delay stage controls the pump's arrival time at the sample. Pump and probe paths are recombined using a non-polarizing beamsplitter.

In both cases, pump and probe are focused and collected using microscope objectives (Mitutoyo 10$\times$ Plan Apo, NA=0.28), and the probe can be directed either for a spectrometer (OceanOptics HR 4000) or a camera (Thorlabs Zelux CS165MU).

\backmatter

\bmhead{Supplementary information}

\begin{itemize}
    \item Full description of the experimental setup used for CW and pulsed laser spectroscopy and imaging. 
    \item Fitting of WS$_2$ permittivity as function of CW pump intensity. 
    \item Analysis of hysteresis on the CW modulation response. 
    \item Longer duration pump-probe scan and analysis of exciton dynamics under different pump fluences.
    \item Pump probe spectroscopy using a 405 nm pump.
    \item CW and pulsed optical pumping of a bare Si$_3$N$_4$ metasurface.
    \\
\end{itemize}

\bmhead{Acknowledgements}

This work was funded by the Open Technology Program of the Dutch National Science Foundation (NWO), grant number 19486. JvdG is also supported by a Vidi grant (VI.Vidi.203.027) from The Netherlands Organization for Scientific Research (NWO), as well as an European Research Council Starting Grant under grant agreement No. 101116984. AA acknowledges support from the Simons Foundation and the Office of Naval Research. We gratefully acknowledge T. Hoekstra for the language screening.

\section*{Declarations}
\bmhead{Author contributions}

B.S.D., M.G., A.C., A.P., A.A and J.v.d.G. conceived the concepts behind this research. B.S.D. and A.C. fabricated the sample. B.S.D., R.T. and M.G. performed the optical measurements. B.S.D. performed the AFM measurements. B.S.D. and R.T. performed the data analysis and calculations. All authors contributed to writing the manuscript.

\textbf{Data availability} - A full replication package including all data and analysis scripts will be made freely available upon publication.

\bibliography{sn-bibliography}% common bib file

\end{document}